\pdfoutput=1

\documentclass[conference]{IEEEtran}

\usepackage[pdftex]{graphicx}
\usepackage{amsmath}
\usepackage{eqparbox}
\usepackage[labelformat=empty]{caption}
\usepackage{hyperref}
\usepackage{tabularx,booktabs}
\usepackage{multirow}
\usepackage{silence}
\hypersetup{%
  pdfborder = {0 0 0},
  bookmarksdepth = section,
  citecolor = DarkGreen,
  linkcolor = DarkBlue,
  urlcolor = DarkGreen,
}%

\begin{document}

\title{\LARGE VEDAR: Accountable Behavioural Change Detection}

\author{\IEEEauthorblockN{Amit Kumar\IEEEauthorrefmark{1},
Tanya Ahuja\IEEEauthorrefmark{2}, Rajesh Kumar Madabhattula\IEEEauthorrefmark{3}, Murali Kante\IEEEauthorrefmark{4},
Srinivasa Rao Aravilli\IEEEauthorrefmark{5}}
\IEEEauthorblockA{
Email: \IEEEauthorrefmark{1}amitkku@cisco.com,
\IEEEauthorrefmark{2}tanahuja@cisco.com,
\IEEEauthorrefmark{3}rmadabha@cisco.com,
\IEEEauthorrefmark{4}mukante@cisco.com,
\IEEEauthorrefmark{5}saravill@cisco.com}}


\maketitle
\thispagestyle{plain}
\pagestyle{plain}


\begin{abstract}
With exponential increase in the availability of telemetry / streaming / real time data, understanding contextual behavior changes is a vital functionality in order to deliver unrivalled customer experience and build high performance and high availability systems. Real time behavior change detection finds a use case in number of domains such as social networks, network traffic monitoring, ad exchange metrics etc. In streaming data, behavior change is an implausible observation that does not fit in with the distribution of rest of the data. A timely and precise revelation of such behavior changes can give us substantial information about the system in critical situations which can be a driving factor for vital decisions. Detecting behavior changes in streaming fashion is a difficult task as the system needs to process high speed real time data and continuously learn from data along with detecting anomalies in a single pass of data. In this paper we introduce a novel algorithm called Accountable Behavior Change Detection (VEDAR) which can detect and elucidate the behavior changes in real time and operates in a fashion similar to human perception. We have bench marked our algorithm on open source anomaly detection datasets. We have bench marked our algorithm by comparing its performance on open source anomaly datasets against industry standard algorithms like Numenta HTM and Twitter AdVec (SH-ESD). Our algorithm outperforms above mentioned algorithms for behaviour change detection, efficacy is given in section V.
\end{abstract}



\section{Introduction}

\textbf{A}CCOUNTABLE behavior change detection is a fundamental problem in data mining which has been well explored in the past few decades. But with an exponential increase in the availability of telemetry data, we are now in a need of a behavioral change detection system that can work well in dynamic and streaming environment. The expansion of Internet of Things (IOT) has added innumerable sources of Big Data into the Data Management landscape. Cloud-Servers, smart phones, sensors on machines, all generate huge amount of real time and continuously changing data for IOT. This leaves us to reconsider the problem of behavior change detection in streaming fashion.

Real time accountable behavioral change detection is gaining practical and significant usage across many industries like analysis of network traffic, monitoring real time tweets, analyzing trends in online shopping, monitoring ad exchange data, monitoring system resource utilization in cloud servers and many more. The significance of behavior change detection is due to the fact that abnormalities in data often lead to significant, often critical, actionable information.

We refer to change in behavior as an observation or a group of observations that are significantly different from rest of the data or the patterns that do not conform to the notion of normal behavior. Abnormal behavior in data can either be spatial or temporal. In spatial behavior change, an individual data point is significantly different from rest of the data, independent of its location in the data stream. The spikes in Fig. 3. represent instances of spatial behavior change. Temporal anomalies occur when a data point is abnormal only in specific temporal context, eg. the first anomalous point (marked in green) in Fig. 4. Such deviations may be caused by a positive factor like increased traffic to a site or a negative reason like CPU utilization exceeding the limits. But in either case such changes can lead to actionable intel.

Behavioral change detection has traditionally been handled using rule-based techniques applied to static data in batches. But with the number of scenarios out-growing in streaming data, such techniques are difficult to scale. Moreover, even with huge amount of streaming data available, the chances of such data being labelled is very rare which moves the traditional classification algorithms out of scope. The behavioral change detection systems face two major challenges in streaming environment. Firstly, streaming analytics requires models that can learn continuously in real time, without storing the entire data, have low operational complexity and are fully automated. Secondly, the definition of anomaly continuously changes as systems evolve and behaviors change. With streaming data being dynamic, the model needs to re-train quickly and adjust to the changing data distribution in real time.

In this paper we present a novel algorithm that detects changes in data behavior in real time. The algorithm takes into consideration, the factors such as seasonal repetition of values and trend in data while determining behavior changes. This system can be applicable to streaming data with different scales like number of bytes written to disk/sec and CPU utilization percentage as it employs a data normalization module. It uses data-driven, dynamic rules to detect abnormalities that can quickly detect sudden changes and also adjust to long term changes in statistics of data. The resulting system is computationally efficient and does not require any prior parameter adjustment. This algorithm provides an edge over existing systems as it adds the feature of accountability. We will explain all these layers in greater depth in section 3. An end to end deployment architecture of VEDAR is shown in Figure 1. 

\begin{figure*}[ht!] 
\centering
\includegraphics[width=5in]{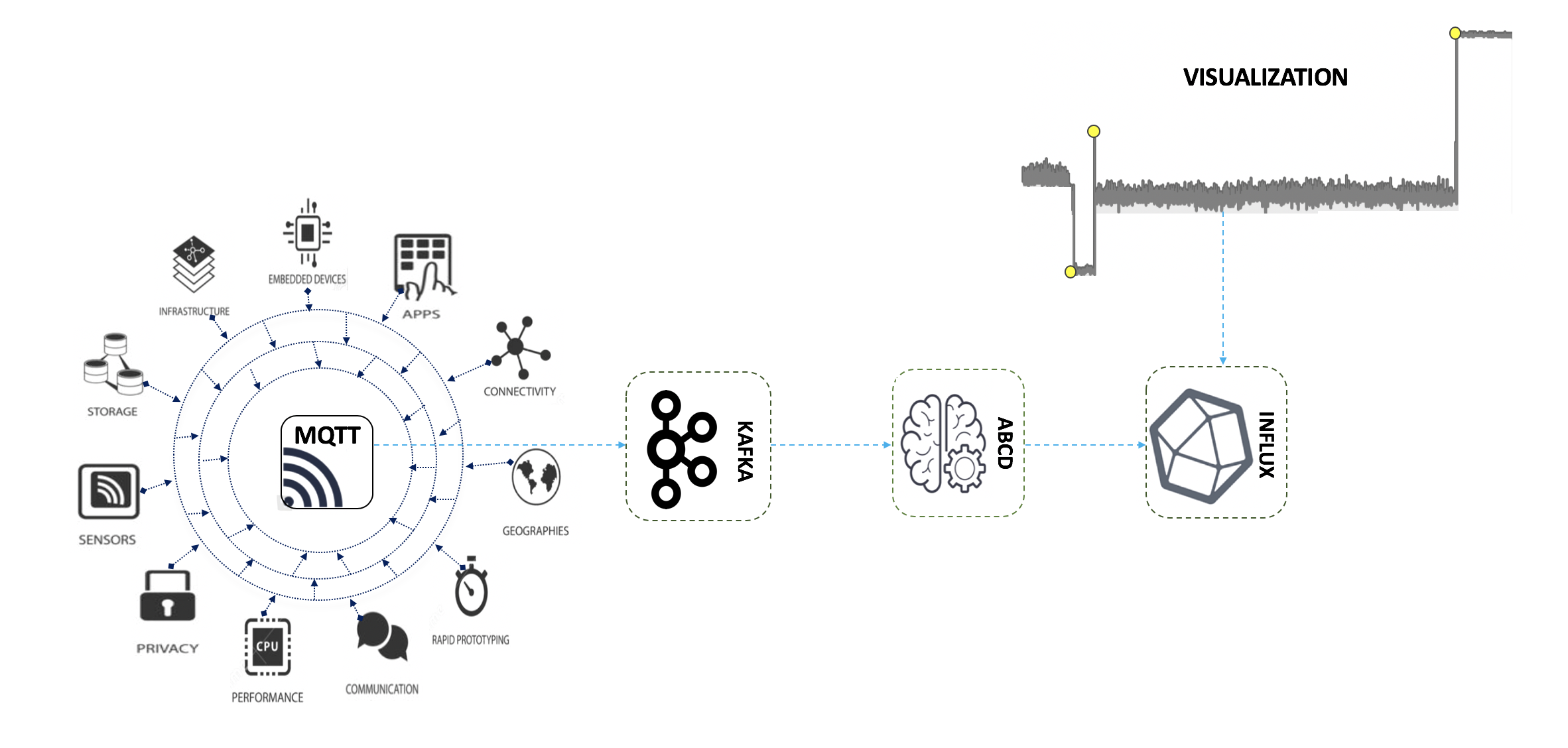}
\caption{Fig. 1. The figure shows Accountable Behavior Change detection deployment architecture in Cloud, IOT, Networking, Operations, Management and Services.}
\label{picture}
\end{figure*}


\section{Related Work}

Behavioral change detection in telemetry/ streaming data has been a well-explored research area in past few decades. Some classical statistical algorithms like setting threshold, moving average and moving median were extensively used for batch anomaly detection. Some other techniques like clustering and exponentially weighted moving average EWMA \cite{S.W.Roberts} have also been studied as a solution for behavioral change detection in streaming data. But these techniques can only be used to detect spatial abnormalities. These techniques are capable of detecting any abrupt change in the process behavior but they fail to adjust to gradual shift in distribution of data. Moreover, these techniques are incapable of taking factors like seasonality and trend into consideration. 

A well-known technique for detecting spatial as well as temporal anomalies in streaming data is change point detection technique \cite{Adams}. These techniques are fast and are also applicable to multivariate time series data. The only limitation of these algorithms is that their performance is sensitive to hyper parameters such as window size and thresholds. Due to this they might result in more false positives. These methods also do not take seasonality into account.

Another algorithm for detecting temporal behavior changes is ARIMA \cite{Bianco}. It is better than above mentioned algorithms as it is a combination of both auto-regressive and moving-average. Seasonal-ARIMA \cite{Adhistya} can also detect seasonality patterns of fixed periods like weekly, daily seasonality. But it fails to re-adjust to shifts in the seasonal patterns in real time. 

Deep learning algorithms like LSTM autoencoders \cite{lstm} have also been explored for detecting behavior changes. It reconstructs models where some form of reconstruction error is used as a measure of a behaviour change. Deep learning models have to be re-trained frequently in order to stay updated with new data. Also, they require huge amount of data for training purpose.

Numenta introduced a machine learning algorithm derived by neuroscience called Hierarchical Temporal Memory \cite{Subutai} which models the spatial and temporal patterns in streaming data. HTM performs better than above mentioned algorithms as it comprehends the seasonal and trend factors in the streaming data and also adjusts with changing data statistics. But HTM does not provide any accountability for detected abnormalities and the cause and type of detected anomalies cannot be interpreted.

Twitter also released an open-source algorithm for detecting both spatial and temporal anomalies in streaming analytics called TwitterAdVec (SH-ESD) \cite{sh-esd}. Although this algorithm models the seasonal patterns, just like HTM, it does not provide any accountability for the detected anomalies. Moreover, the precision of SH-ESD is less as compared to VEDAR because of large number of false positives. We have compared the results of our algorithm with HTM and Twitter Anomaly Detection Algorithm in the comparison section.


\section{VEDAR FOR REAL TIME ACCOUNTABLE BEHAVIORAL CHANGE DETECTION}
For the purpose of this paper, we define behavior change as an observation or a group of observations that deviate significantly from underlying distribution of rest of the data. In this paper we focus primarily on 3 types of behavior changes: (i) Seasonality interruption change: when the observation deviates from expected seasonal value, illustrated in Figure 2, (ii) Erratic change: an abrupt transient change that is short lived, as shown in Figure 3, (iii) Linear change: when observations gradually proceed towards abnormal behavior, depicted in first anomaly point in Figure 4. For the task of change detection, it is vital to detect erratic changes, initially alert in case of seasonality interruption and have the model must quickly re-adjust to such change and also alert in case the signal is gradually approaching towards abnormal behavior.

\begin{figure*}[ht!] 
\centering
\includegraphics[width=6.5in]{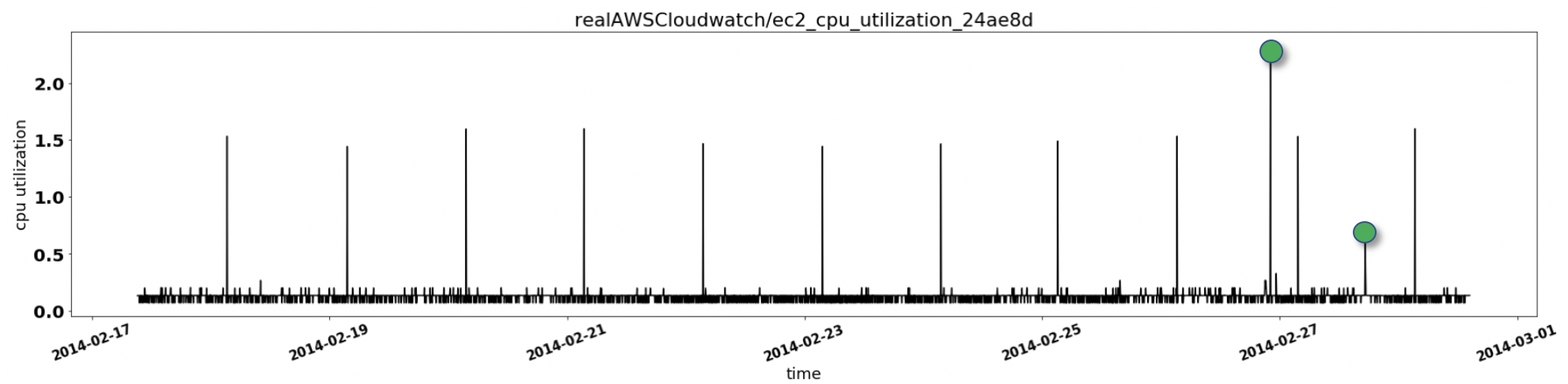}
\caption{Fig. 2. The figure is a real-world CPU utilization data from AWS Cloud Watch collected over 5 minutes interval. Data contains periodic spikes occurring at a frequency of 1 day. The anomalous points are caused by interruption of seasonality.}
\label{picture}
\end{figure*}

\begin{figure*}[ht!] 
\centering
\includegraphics[width=6.5in]{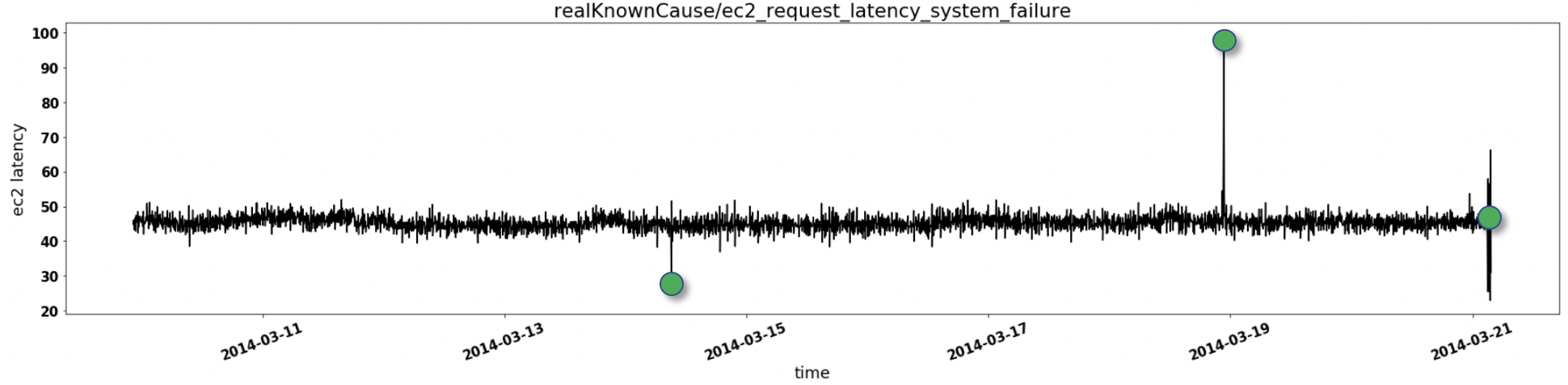}
\caption{Fig. 3. The figure is a real-world ec2-request-latency-system-failure data collected from AWS Cloud Watch servers over 5 minutes interval. Anomalous points depict an erratic change in system behavior.}
\label{picture}
\end{figure*}

\begin{figure*}[ht!] 
\centering
\includegraphics[width=6.5in]{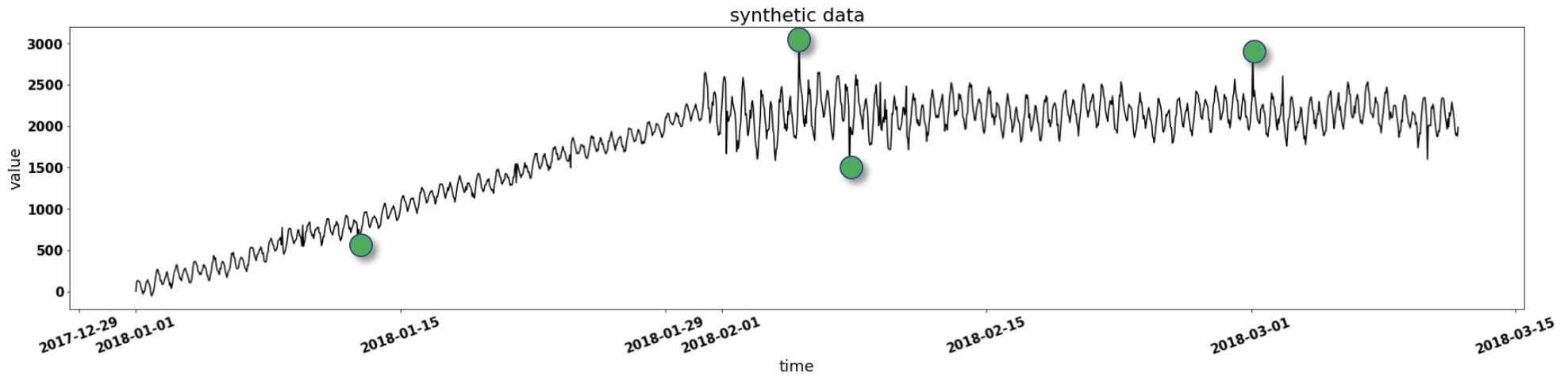}
\caption{Fig. 4. The figure is an artificial data depicting observations linearly trending towards the upper limit followed by periodic data with random spikes. First anomaly point captures the linear trend and second and third anomalies depict erratic changes in signal value.}
\label{picture}
\end{figure*}

In order to identify change in signal behavior, VEDAR scales the input signal, redirects the normalized signal to analyze any non-stationary factors like trend and seasonality. If either of these are present, the signal is made stationary by subtracting out trend and seasonal components. The residual is further smoothened out using exponential and linear quadratic smoothening techniques. The smoothened signal is further utilized by Non-Parametric Estimation module in order to generate likelihood of the observation belonging to the underlying distribution. The system detects an event stream as an anomaly based on empirical rules. Figure 5 illustrates a block diagram of our algorithm. 

We explain each layer of VEDAR in following subsections. 

\begin{figure}[ht!] 
\centering
\includegraphics[width=3.45in]{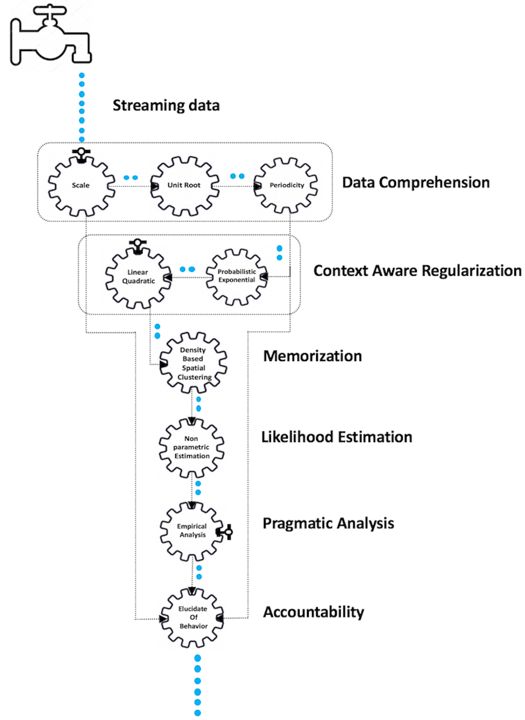}
\caption{Fig. 5. Flow diagram of VEDAR.}
\label{picture}
\end{figure}

\subsection{Layer 1 - Data Comprehension}
In the first layer, system understands the nature of streaming data by extracting features like trend, seasonality and scale of data. According to the observed characteristics of data, the algorithm processes the input signal in order to perform equally well on datasets with different characteristics.
This layer consists of three sub-modules namely, data scaling, trend and seasonality extraction. These sub-layers are explained in below subsections.

\subsubsection{Data Scaling}
Scaling of data is a feature in order to deal with data sets of varying scales and also to detect anomalies at desired granularity. For instance, if the input data is in bytes, we are able to detect behavior changes at the scale of Gigabytes by scaling the entire data to Gigabytes. This provides the user with the flexibility to choose the granularity of detected behavior changes.

VEDAR is capable of detecting behavior changes in input signals with different scales like number of bytes written to disk/sec which can range from 0 to hundreds of Gigabytes as well as percentage CPU utilization which lies between 0 to 100. In order to operate equally well on such different scales, the system internally maintains a data-driven scaling/normalization factor. The input signal is divided by this normalization factor to scale it. The value of this scaling factor can be controlled in 2 ways: (1) Users can explicitly provide a value for the scaling factor based on the granularity of data and percentage of data expected as anomalies. For instance, if the streaming data denotes the number of bytes written to disk/sec and the user expects the fluctuations of Gigabytes/sec to be detected as behavior changes then a scaling factor of 1024 or 1024*1024 is appropriate. This converts entire data into Gigabytes scale and behavior changes are detected at this scale. (2) Alternatively, VEDAR automatically determines the scaling factor by observing set of data points. The normalization factor is dynamic and is computed frequently by the system, thus enabling us to detect any behavior change.

\subsubsection{Seasonality and Trend Extraction}
As already mentioned, some data streams are seasonal in nature i.e., the system behavior changes periodically resulting in periodic fluctuations. It is essential to detect any periodicity present in data in order to avoid false alarms in case of periodic behavior changes. In absence of seasonality and trend extraction module, the algorithm would raise an alarm at every seasonal fluctuation which leads to huge number of false positives.

Since the input signal might be non-stationary, VEDAR first extracts the trend and seasonal components from the signal in order to make it stationary. As already mentioned, in a streaming data some of the observations or behavior changes occur quasi-periodically. That is, values in same range occur at a similar time every hour/day/week/year. In order to calculate the periodicity of data, our algorithm uses YIN \cite{prefix}, an auto-correlation \cite{YIN} based method which is explained in greater depth in. In YIN, the difference function is calculated as follows,

\begin{equation}
\small
D^{'}(\rho,t_j) = \frac{D(\rho, t_j)}{\frac{1}{p} \sum_{k=1}^{\rho}D(k,t_j)}
\end{equation}
where, 
\begin{equation}
\small
D(\rho,t_j) = \sum_{k=1}^{h-\rho}(C_{\xi}^{k}(t_j) - C_{\xi}^{k+\rho}(t_j) )^2
\end{equation}

The terms of above function are explained clearly in section 3.4.2 of \cite{prefix}. On calculating correlation values $D^{'}$ for each possible value of $\rho$, we detect the lowest value (troughs) in the de-trended correlation values. The frequency of periodicity is the value $\rho$, whose correlation corresponds to the lowest trough. Let us denote this value by $\rho^{'}$.

The scaled input signal is first de-seasonalized if any periodicity is detected in the previous observations. In order to de-seasonalize the input signal, a window of size w is taken around the previous seasonal data point (observation occurring exactly $\rho^{'}$  points prior to current point in the data stream) and current data point is subtracted from the most similar observation in the seasonal window.

VEDAR outperforms a number of available algorithms by quickly adapting to changes in periodicity of data. As illustrated in Figure 6, seasonality of data varies with time. VEDAR adapts to such change in periodicity by continuously updating the value of $\rho^{'}$. At the end of each day, periodicity value $\rho^{'}$ is re-calculated based on past 2 weeks of data.

\begin{figure*}[ht!] 
\centering
\includegraphics[width=6.5in]{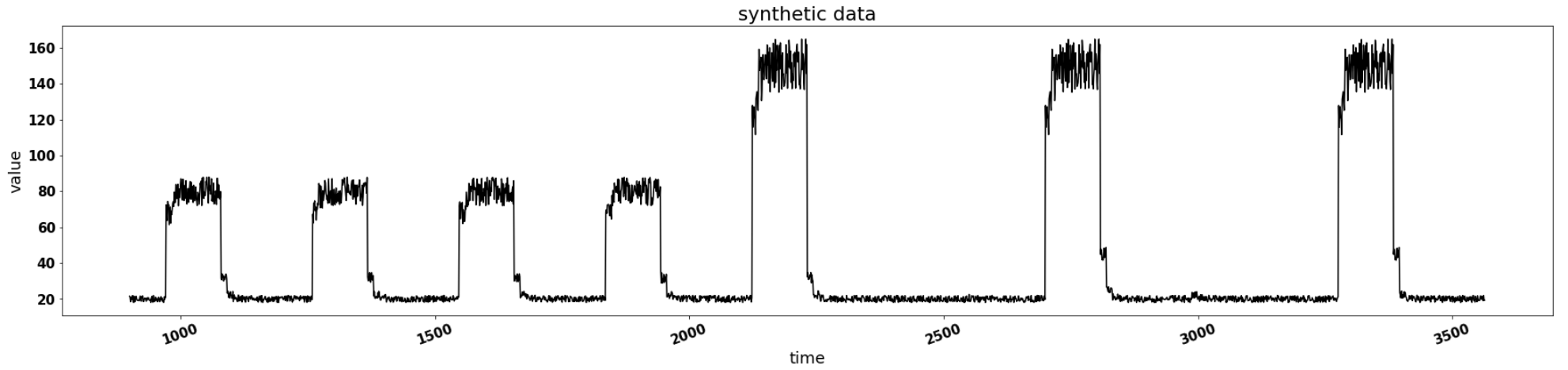}
\caption{Fig. 6. The figure is an artificial data generated at a frequency of 5 minutes. The data illustrates periodic steps after every 24 hours. At certain point, periodicity shift from 24 hours to 48 hours is evident in the graph.}
\label{picture}
\end{figure*}

\subsection{Layer 2 - Context Aware Regularization}
Despite of the presence of seasonality in data, the odds of such seasonality being perfect are very rare i.e., periodic fluctuations are seldom of equal magnitudes. The residual signal obtained after seasonality and trend extraction incorporates a lot of white noise which is caused as a consequence of disproportionate periodic behavior changes. 

The presence of white noise in data leads to high number of false positives as any fluctuation caused as a consequence of this white noise is detected by the algorithm as behavior change. In order to eliminate white noise from data we perform a context aware smoothening of the residual signal. Figure 7(a) shows the synthetic signal data and Figure 7(b) shows the de-seasonalized signal data containing white noise.

\begin{figure*}[ht!] 
\centering
\includegraphics[width=6.5in]{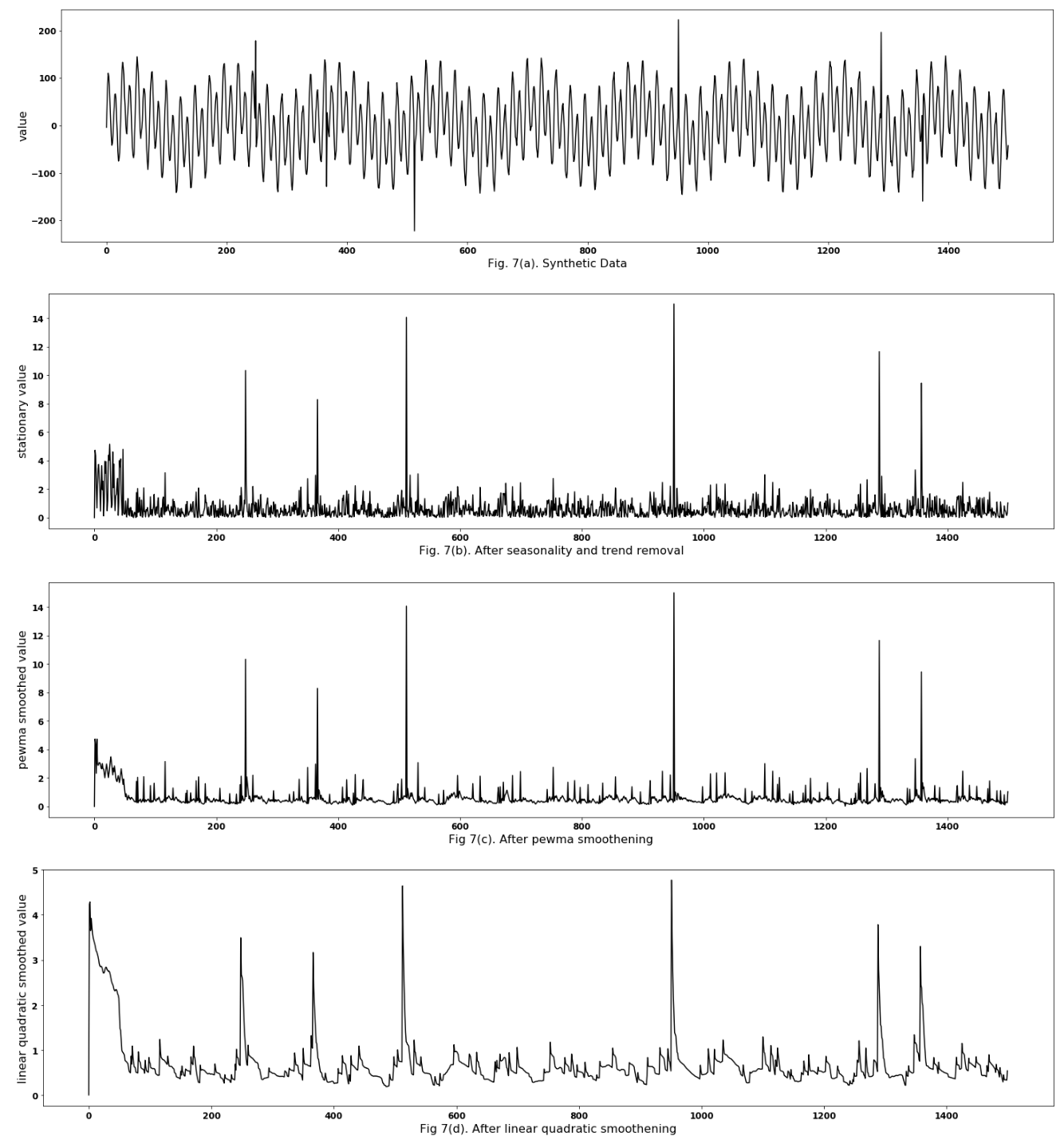}
\caption{Fig. 7. The figure contains 4 graphs. First graph shows a synthetic data generated at a frequency of 5 minutes. Data contains periodic peaks and troughs after every 224 data points. Second graph displays the de-seasonalized and de-trended signal. Data in Graph 2 possesses a lot of white noise. Third graph shows smoothened signal obtained after applying probabilistic exponential smoothening. It levels out smaller variations. Fourth graph contains the residual signal obtained by applying Linear Quadratic smoothening on the previous signal.}
\label{picture}
\end{figure*}

VEDAR performs two levels of smoothening in order to eliminate white noise from the signal while keeping the behavior changes intact. The primary smoothening step which evens out very small fluctuations is probabilistic exponential smoothening. Once the smaller variations are handled, second level of smoothening is applied which tackles any persisting fluctuations while leaving significant behavior changes unaffected. This layer is called linear quadratic smoothening. The necessity for two levels of smoothening arises in order to perform constrained smoothening. The following subsections explain the smoothening techniques used in VEDAR.

\subsubsection{Probabilistic Exponential Smoothening}
This is an advanced version of Exponential smoothening \cite{S.W.Roberts} for time series data. Exponential moving average computes local mean \textit{$\mu_t$} of a time series \textit{${X_t}$} by applying exponentially decreasing weight factors to the past observations. Weighting factor \textit{$\alpha$} determines the amount of weight given to historic values. 

\begin{equation}
\small
    \mu_t = \alpha\mu_{t-1} + (1 - \alpha)X_t
\end{equation}

Due to fixed weighting factor \textit{$\alpha$}, moving mean \textit{$\mu_t$} is highly susceptible to any abrupt behavior. 

Probabilistic Exponential Smoothening which is a modified version of PEWMA \cite{pewma} adjusts the weighting parameter \textit{$\alpha$} based on probability of current observation \textit{${X_t}$}. 
\begin{equation}
\small
    \mu_t = \alpha(1-\beta P_t)\mu_{t-1} + (1-\alpha(1-\beta P_t))X_t
\end{equation}

Where \textit{$P_t$} is the probability of \textit{${X_t}$} based on underlying data distribution and \textit{$\beta$} is the weightage given to \textit{$P_t$}. For the purpose of smoothening, any observation with $|X_{t}$ - $ \mu_{t} | < 3\alpha_t $ is substituted with whereas the observations that are greater than $3\alpha_t$ away from $ \mu_{t}$ are kept unaffected. PEWMA smoothened signal is displayed in Figure 7(c).

\subsubsection{Linear Quadratic Smoothening}
Linear quadratic smoothening \cite{kalman} is a widely used technique for signal smoothening. It begins by predicting the next value based on previous observations. Based on the deviation of this predicted value from the actual data point, the predicted value is either directly utilized as smoothened value or the system is trained with actual value to reduce the error term. The signal obtained after applying Linear Quadratic Smoothening is shown in Figure 7(d).

Linear Quadratic Smoothening consists of two steps: (1) Prediction: Linear Quadratic Smoothening produces estimates of the current state variables, along with their uncertainties.

\begin{equation}
\small
    X^{-} = FX
\end{equation}
\begin{equation}
\small
    P^{-} = FPF^{T}
\end{equation}

where \textit{$X^-$} is the predicted value, \textit{X} is the value predicted in the previous step. \textit{F} is state transition matrix which computes \textit{$X^-$} given \textit{X} (For our use case, value of \textit{F} is taken as 1). \textit{$P^-$} is the predicted process co-variance matrix and \textit{P} is the process co-variance matrix predicted in the previous step. These are used to model the variance of the system.
(2) Error correction: Once the outcome of the next measurement (necessarily corrupted with some amount of error, including random noise) is observed, these estimates are updated using a weighted average, with more weight being given to estimates with higher certainty.

\begin{equation}
\small
    S = HP^{-}H^{T}
\end{equation}
\begin{equation}
\small
    K = P^{-}H^{T}S^{-1}
\end{equation}
\begin{equation}
\small
    Y = Z - HX^{-}
\end{equation}
\begin{equation}
\small
    X = X^{-} + KY
\end{equation}
\begin{equation}
\small
    P = (I - KH)P^{-}
\end{equation}

where \textit{H} is measurement function used to transform state variables into measurement space and \textit{S} is the transformation of Process co-variance matrix in measurement space. \textit{K} is Kalman gain is the ratio between uncertainty in prediction and uncertainty in measurement. \textit{Y} is the deviation of actual data point \textit{Z} from the predicted value \textit{$X^-$}. \textit{X} is the updated estimate value which is weighed sum of the actual prediction \textit{$X^-$} and the residual \textit{Y}. Kalman gain \textit{K} determines the weight given to the residual value \textit{Y} in order to update the predicted value. State co-variance matrix \textit{P} is also updated using Kalman gain and measurement function \textit{H}.

Details of Linear quadratic smoothening technique are available in \cite{kalman}.

\subsection{Layer 3 – Memorization}
In order to distinguish behavior changes from rest of the observations, some technique is required to segregate and memorize the past behavior. The absence of context knowledge leads to high number of false alarms as the algorithm fails to learn from previously available data.

Distribution generated over the past behavior is used by the algorithm to determine that any given observation is anomalous or not. Generating a distribution over entire historic data is a tedious job as it would increase the memory requirement along with slowing down the system. In order to resolve this issue, VEDAR samples data points from historic data.

VEDAR uses Density based Spatial Clustering technique (DBSCAN) \cite{dbscan} to segregate normal and anomalous behaviors. Samples from these clusters are then used to generate distribution for the data points which is used by the subsequent layer. Since streaming data can be highly dynamic, it leads to the formation of unspecified number of clusters with different densities. As DBSCAN can perform well on such datasets, it seemed to fit the use case.

\subsection{Layer 4 - Likelihood Estimation}
In order to determine the probability of any given data point belonging to the underlying signal distribution, we first need to establish an underlying data distribution. VEDAR employs KDE \cite{kde} for this purpose. 

Kernel Density Estimation is a non-parametric estimation technique which models combinations of distributions. KDE estimates the probability density function of random variables. It is applied at two different steps in VEDAR. (1) Each cluster of historic data generated by DBSCAN is supposed to have an underlying distribution. These distributions are modeled using KDE. (2) Once the distributions are identified, samples are generated from them using random sampling. The sampled data points from historic data along with current data buffer are used to train KDE for determining the likelihood of any given observation.  

Given an input data point, KDE determines its likelihood by applying Kernel function. For the purpose of behavior change detection VEDAR uses Gaussian kernel.

\begin{equation}
\small
    P_k(y) = \sum_{i=1}^{N} K((y-x_{i})/h)
\end{equation}
where \textit{K} is Gaussian kernel and \textit{h} is the bandwidth which is a hyper-parameter controlling the spread of distribution over data. \textit{N} is number of data points. \textit{y} is data point for which we want to compute likelihood.

\subsection{Layer 5 - Pragmatic Analysis}
Empirical rule is often used in statistics for the purpose of forecasting. VEDAR use empirical rule on the likelihood of data points to determine if given data point is anomalous or not. We apply two set of rules for analyzing behavior of streaming data.

(1) Empirical rule or Three Sigma Rule – If the likelihood of current data point deviates from the previous data point’s likelihood by more than three sigma value, then the current observation is considered to be a significant behavior change. This rule fails to address the issue of linearly dropping probabilities.

(2) In case of observations gradually deviating from the normal behavior, the likelihood of each upcoming observation approaches more and more towards 0. In this case, the behavior change can not be identified with empirical rule as the likelihoods of both of the consecutive observations is exponentially small ($10^{-10}$ scale and below)and so is the likelihood difference. In order to handle the case where the observations are linearly deviating from the underlying data distribution, we monitor the scale change in the likelihood of a set of data points rather than monitoring the actual difference of likelihood.

\subsection{Layer 6 - Accountability}
Accountability is the layer of VEDAR which makes it superior to rest of the algorithms available for behavior change detection. In this layer, the algorithm rationalizes the cause of such behavior change based on the type of anomaly occurred. Whenever a behavior change happens, the algorithms raises an alarm to the user which contains the following information: actual value, expected value and the type of behavior change detected. Based on the type of behavior change, user can take appropriate actions.

We have explained the three types of behavior changes detected by VEDAR earlier in Section 3. These are seasonality interruption change, erratic change and linear changes. Please refer to Figures 2, 3 and 4 for visualization of these 3 categories.

Any behavior changes detected in signal containing seasonality are identified as seasonality interruption changes as the system behaves differently from the expected seasonal behavior.

In the datasets which do not contain periodicity, any sudden probability drop of an observation is denoted as erratic change whereas the linearly dropping scale of likelihood is designated as linear change.

\begin{table*}[]
\centering
\begin{tabular}{ |c||c|c|c|c|c|c|c| }
 \hline
 \textbf{DataSet} & \textbf{Algorithm} & \textbf{True Positive} & \textbf{False Positive} & \textbf{False Negative} & \textbf{Precision} & \textbf{Recall} & \textbf{F1-Score}\\
 \hline
\multirow{3}{15em}{realKnownCause/ec2\_request\_latency \newline \_system\_failure} & HTM & 3 & 9 & 0 & 0.25 & 1 & 0.4 \\ 
& TwitterAdVec & 3 & 5 & 0 & 0.38 & 1 & 0.55 \\ 
& VEDAR & 3 & 1 & 0 & 0.75 & 1 & 0.86\\

\hline
\multirow{3}{15em}{realAWSCloudwatch/rds\_cpu \newline \_utilization\_e47b3b} &HTM & 2 & 2 & 0 & 0.5 & 1 & 0.67\\
&TwitterAdVec & 1 & 7 & 1 & 0.13 & 0.5 & 0.2\\
&VEDAR & 2 & 0 & 0 & 1 & 1 & 1\\

\hline
\multirow{3}{15em}{realTraffic/occupancy\_6005} &HTM & 1 & 1 & 0 & 0.5 & 1 & 0.67\\
& TwitterAdVec & 1 & 3 & 0 & 0.25 & 1 & 0.4\\
& VEDAR & 1 & 0 & 0 & 1 & 1 & 1\\

\hline
\multirow{3}{15em}{realTweets/Twitter\_volume\_FB} & HTM & 2 & 4 & 0 & 0.33 & 1 & 0.5\\
& TwitterAdVec & 2 & 26 & 0 & 0.07 & 1 & 0.13\\
& VEDAR & 2 & 1 & 0 & 0.67 & 1 & 0.8\\

\hline
\multirow{3}{15em}{realAdExchange/exchange-4\_cpm\_results} & HTM & 3 & 4 & 0 & 0.43 & 1 & 0.6\\
& TwitterAdVec & 2 & 1 & 1 & 0.67 & 0.67 & 0.67\\
& VEDAR & 3 & 1 & 0 & 0.75 & 1 & 0.86\\
\hline
\end{tabular}
\caption{Table 1: Performance results of HTM, Twitter Anomaly De- tection and VEDAR on NAB datasets.}
\end{table*}

\section{Results}
We have tested our algorithm on the Numenta Anomaly Benchmark Dataset. The dataset comprises of real time anomaly detection data collected over five different domains. In this section, we present the results of VEDAR on one dataset from each category. Each of the Figures 8 to 12 represent a real-world telemetry data belonging to each of the categories. In each of the following graphs, the actual behavior changes, as provided by Numenta are marked by green dots while the behavior changes detected by VEDAR algorithm are marked by yellow dots. The overlapped green and yellow dots represent the correctly detected behavior changes / true positives while the individual green and yellow dots represent the false negatives and false positives respectively.

We have compared the results of our Algorithm with Numenta HTM and TwitterAdVec (SH-ESD) algorithm as well. The accuracy metrics used for this comparison are number of True Positives, False Positives, False Negatives, precision, recall and F1-Score. 

The final comparison of all the three algorithms over these five datasets is shown in Table 1 in the comparison section. 

Figure 8 shows the real CPU utilization data collected over AWS servers by Amazon Cloud Watch Service. The observations collected over a frequency of 5 minutes range between 0 to 100 as the data is percentage of CPU utilized. Other data sets in this category contain AWS server metrics like Network bytes In, Disk write bytes, CPU utilization etc. The data snapshot exhibits 2 significant behavior changes, first one being an erratic spike and the second one is a sudden drift in the distribution of data.

\begin{figure*}[ht!] 
\centering
\includegraphics[width=6.5in]{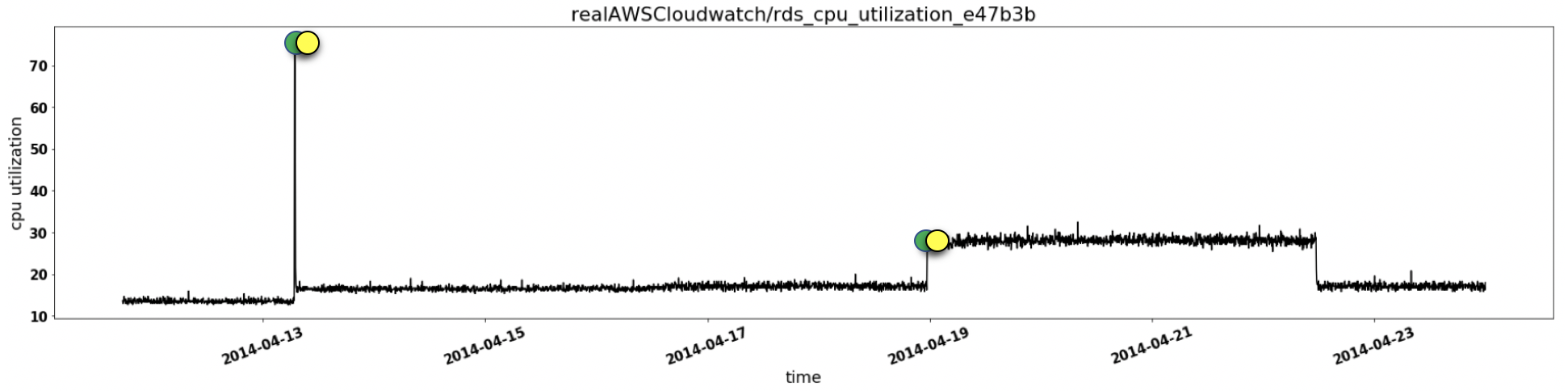}
\caption{Fig. 8. The figure shows real CPU Utilization data collected by Amazon Cloud Watch Service over a frequency of 5 minutes. The data contains two actual significant behavior changes, both of which are detected accurately by VEDAR algorithm as represented by overlapped yellow and green dots.}
\label{picture}
\end{figure*}

Figure 9 illustrates a traffic occupancy dataset from the category of Real Traffic over a frequency of 5 minutes. This category contains real time traffic data from the twin Cities Metro area in Minnesota, collected by the Minnesota Department of Transportation. Metrics captured by the sensors in this domain include occupancy, speed and travel time. The dataset contains periodic repetition of values after every 14 hours. Straight line in the Figure indicates missing data. Behavior of the data deviates from the expected seasonal behavior once in the given snapshot which is accurately detected by VEDAR algorithm as shown in the figure.

\begin{figure*}[ht!] 
\centering
\includegraphics[width=6.5in]{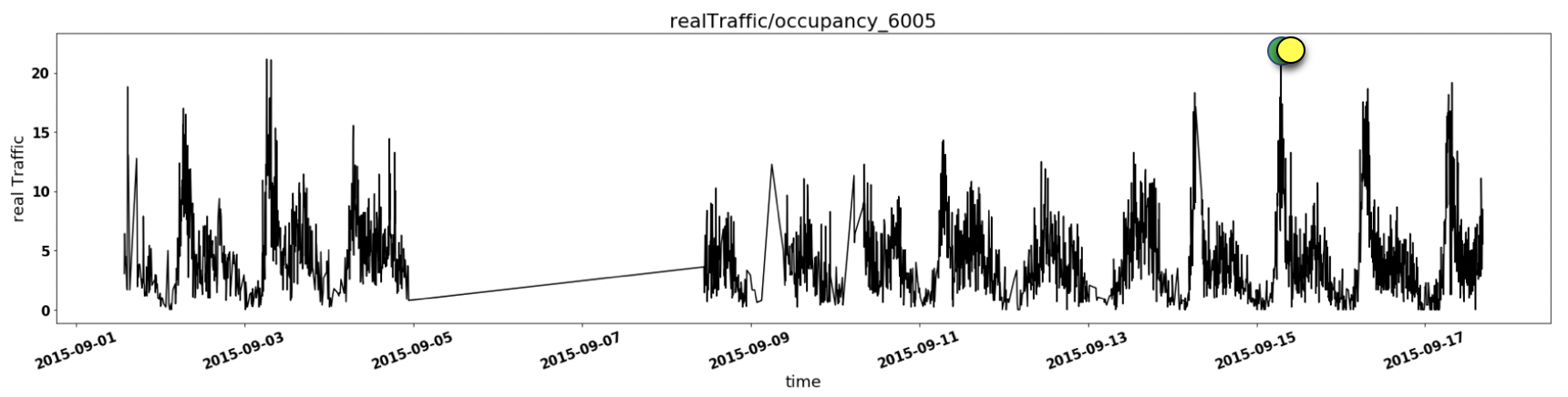}
\caption{Fig. 9. The figure is a real traffic occupancy data collected over a frequency of 5 minutes. The data contains 1 significant behavior change marked by yellow and green overlapped dots. The change in behavior belongs to the category of periodicity interruption as the observed values are significantly larger than expected.}
\label{picture}
\end{figure*}
Figure 10 shows a dataset from real Ad-Exchange category. This dataset captures cost per thousand impressions (CPM) which is a metric for online advertisement clicking rates. The dataset is collected over frequency of 1 hour. The data contains rare erratic spikes which are detected by our algorithm accurately. While VEDAR detected all the actual behavior changes correctly, it also detected 1 behavior change (false positive) which is not mentioned by NAB. 

\begin{figure*}[ht!] 
\centering
\includegraphics[width=6.5in]{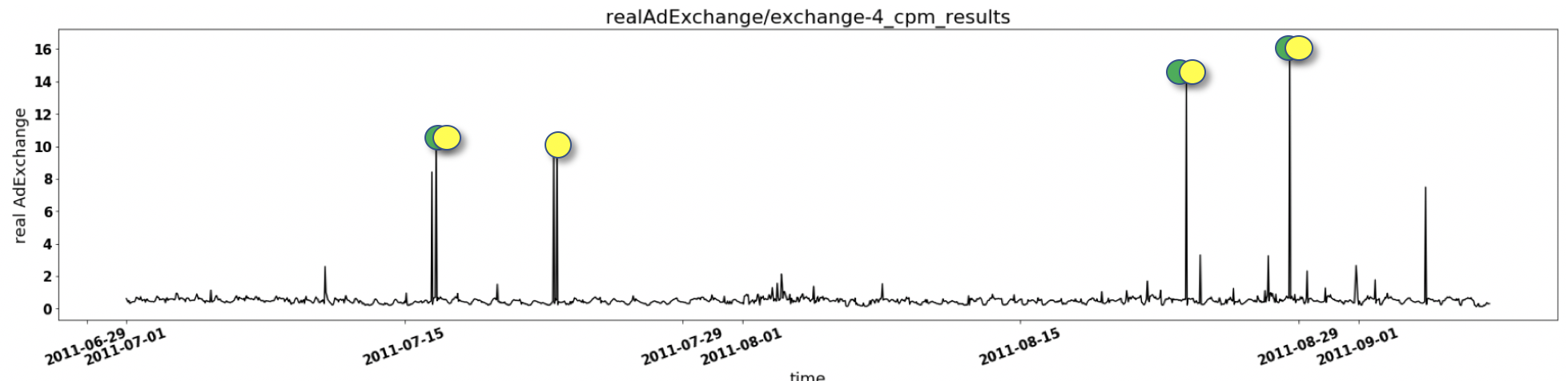}
\caption{Fig. 10. The figure is a real ad exchange dataset which measures cost per thousand impressions. Data contain 4 significant spikes as illustrated in the figure.}
\label{picture}
\end{figure*}
Another category of datasets provided by Numenta is Real Tweets which is a collection of Twitter mentions of large publicly traded companies like Google, Facebook etc. Dataset used for demonstration in this paper contains the number of mentions of Facebook in tweets every 5 minutes. Results of VEDAR on Facebook Real tweets is illustrated in Figure 11. The snapshot of data contains 2 significant behavior changes which are correctly detected by our algorithm, while giving one false positive. Both of the behavior changes are sudden short-lived spikes.

\begin{figure*}[ht!] 
\centering
\includegraphics[width=6.5in]{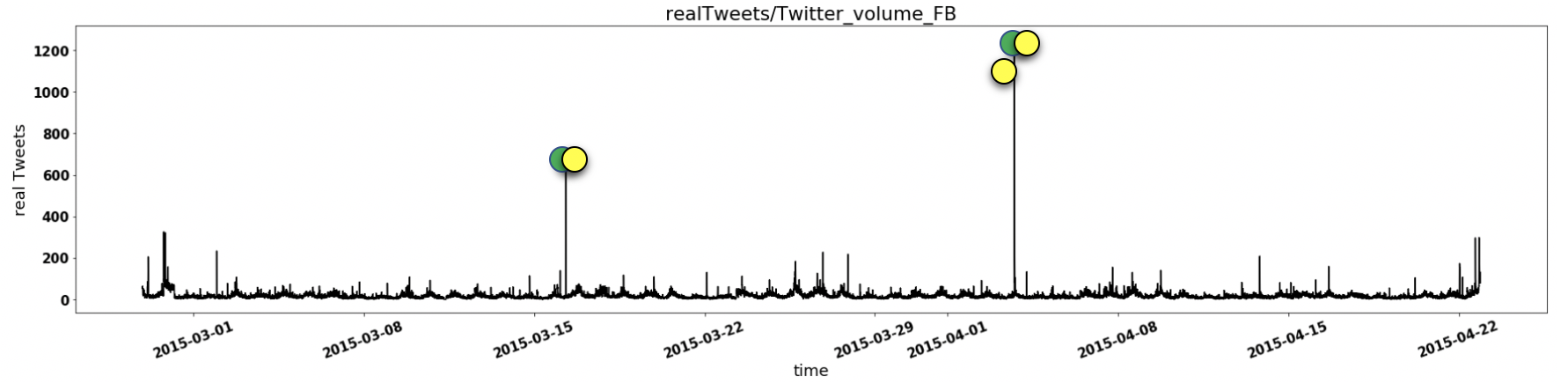}
\caption{Fig. 11. The figure shows tweet mentions of Facebook in 5 minutes intervals. Dataset contains two sudden spikes detected by VEDAR. These behavior changes are displayed by overlapping green and yellow dots.}
\label{picture}
\end{figure*}

The ec2-Request Latency System Failure dataset illustrated in Figure 12 belongs to real Known Cause category. This dataset presents CPU utilization data from a server from Amazon’s east coast data center.  The data snapshot contains 2 erratic spikes and ends with a complete system failure as shown in Figure 12. No periodicity is present in this dataset.

\begin{figure*}[ht!] 
\centering
\includegraphics[width=6.5in]{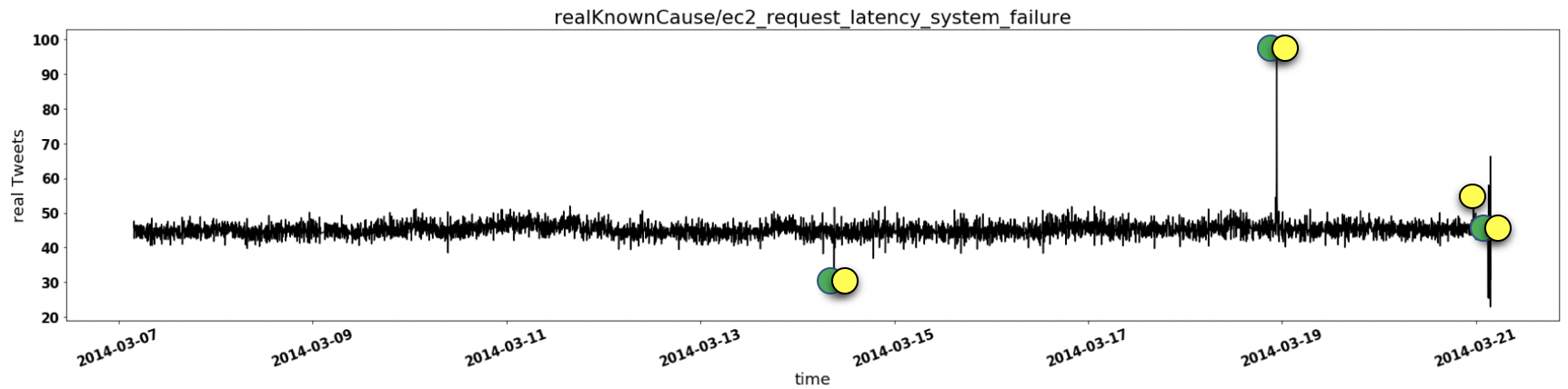}
\caption{Fig. 12. The figure represents the CPU utilization data from one of Amazon’s AWS server collected over a frequency of 5 minutes. As evident in Figure, data contains 4 behavior changes / spikes detected by VEDAR. The last behavior change represents complete system failure. VEDAR detected 3 true behavior changes along with one false alarm depicted in overlapped dots and an individual yellow dot respectively.}
\label{picture}
\end{figure*}


\section{COMPARISON}
This section contains the consolidated results of three algorithms: VEDAR, HTM and  Twitter Anomaly Detection on one dataset from each of the 5 domains from the Anomaly Benchmark datasets provided by \href{https://github.com/numenta/NAB/tree/master/data}{\textit{Numenta}}. Metrics used for comparison of accuracy are number of True positives, false positives, false negatives, precision, recall and F1 score. We have taken HTM and Twitter Anomaly Detection results from  \href{https://github.com/numenta/NAB/tree/master/results/numenta}{\textit{numenta}} and \href{https://github.com/numenta/NAB/tree/master/results/twitterADVec}{\textit{twitterAdVec}} respectively. We have used below formula for HTM and TwitterAdVec to determine anomaly:
\begin{equation}
\small
anomaly\_score >= 1 - \epsilon
\end{equation}
where, $\epsilon = 0.01$.

As it is evident from the results in Table 1, VEDAR gives the least number of false positives on all the datasets while not missing any of the actual behavior changes at the same time (0 false negatives). Precision and recall of VEDAR is higher than both HTM and TwitterAdVec. 

where,
\begin{equation}
\small
    precision = \frac{true positives}{true positives + false positives}
\end{equation}
\begin{equation}
\small
    recall = \frac{true positives}{true positives + false negatives}
\end{equation}
\begin{equation}
\small
    F1-score = 2 * \frac{precision * recall}{precision + recall}
\end{equation}

\section{CONCLUSION}
In this paper we presented a novel algorithm for behaviour change detection that can work well in dynamic real time environment. With the exponential increase in availability of connected real time sensors, behaviour change detection is gaining much importance as an application of Machine learning in IOT.

As the results describe, VEDAR produces best in class results for behaviour change detection on NAB datasets with least number of false positives. While VEDAR is robust to both spatial and temporal anomalies, it also proves to be capable of adapting to changes in generative model of data. It detects both abrupt behaviour changes and slowly growing abnormal behaviour equally well. It 
is computationally efficient and needs no prior tuning of parameters. 

The future extensions for VEDAR include the application of algorithm for multivariate version. VEDAR works as an ensemble of multiple layers where each layer performs a specific operation. In order to further improve the accuracy of the system, exploring other models for individual layers could potentially emerge useful.  
\vspace{1em}
\bibliographystyle{IEEEtran}
\bibliography{IEEEabrv,vedar}

\begin{thebibliography}{10}
\providecommand{\url}[1]{#1}
\csname url@samestyle\endcsname
\providecommand{\newblock}{\relax}
\providecommand{\bibinfo}[2]{#2}
\providecommand{\BIBentrySTDinterwordspacing}{\spaceskip=0pt\relax}
\providecommand{\BIBentryALTinterwordstretchfactor}{4}
\providecommand{\BIBentryALTinterwordspacing}{\spaceskip=\fontdimen2\font plus
\BIBentryALTinterwordstretchfactor\fontdimen3\font minus
  \fontdimen4\font\relax}
\providecommand{\BIBforeignlanguage}[2]{{%
\expandafter\ifx\csname l@#1\endcsname\relax
\typeout{** WARNING: IEEEtran.bst: No hyphenation pattern has been}%
\typeout{** loaded for the language `#1'. Using the pattern for}%
\typeout{** the default language instead.}%
\else
\language=\csname l@#1\endcsname
\fi
#2}}
\providecommand{\BIBdecl}{\relax}
\BIBdecl

\bibitem{S.W.Roberts}
\BIBentryALTinterwordspacing
S.~W. Roberts, ``{Control chart tests based on geometric moving averages},''
  vol.~1, no.~3, pp. 239--250, 1959. [Online]. Available:
  \url{https://www.tandfonline.com/doi/abs/10.1080/00401706.1959.10489860}
\BIBentrySTDinterwordspacing

\bibitem{Adams}
\BIBentryALTinterwordspacing
R.~P. Adams and D.~J.~C. Mackay, ``{Bayesian Online Changepoint Detection},''
  vol.~1, p.~7, 2007. [Online]. Available: \url{http://arxiv.org/abs/0710.3742}
\BIBentrySTDinterwordspacing

\bibitem{Bianco}
\BIBentryALTinterwordspacing
G.~B. M. M. E.~J. Bianco, A.M. and V.~J. Yohai, ``{Outlier detection in
  regression models with ARIMA errors using robust estimates},'' vol.~1, 2001.
  [Online]. Available:
  \url{https://onlinelibrary.wiley.com/doi/abs/10.1002/for.768}
\BIBentrySTDinterwordspacing

\bibitem{Adhistya}
\BIBentryALTinterwordspacing
I.~A.~B. Adhistya Erna~Permanasari, Indriyana~Hidayah, ``{SARIMA(Seasonal
  ARIMA) Implementation on Time Series to Forecast The Number of Malaria
  Incidence},'' vol.~1, 2013. [Online]. Available:
  \url{https://ieeexplore.ieee.org/abstract/document/6676239/authors#authors}
\BIBentrySTDinterwordspacing

\bibitem{lstm}
\BIBentryALTinterwordspacing
G.~A. L. V. P. A. G.~S. Pankaj~Malhotra, Anusha~Ramakrishnan, ``{LSTM-based
  Encoder-Decoder for Multi-sensor Anomaly Detection},'' vol.~2, 2016.
  [Online]. Available: \url{https://arxiv.org/pdf/1607.00148.pdf}
\BIBentrySTDinterwordspacing

\bibitem{Subutai}
\BIBentryALTinterwordspacing
S.~Ahmad and S.~Purdy, ``{Real-Time Anomaly Detection for Streaming
  Analytics},'' 2016. [Online]. Available:
  \url{https://arxiv.org/pdf/1607.02480.pdf}
\BIBentrySTDinterwordspacing

\bibitem{sh-esd}
\BIBentryALTinterwordspacing
O.~S.~V. Jordan~Hochenbaum and A.~Kejariwal, ``{Automatic Anomaly Detection in
  the Cloud Via Statistical Learning},'' vol.~1, 2017. [Online]. Available:
  \url{https://arxiv.org/pdf/1704.07706.pdf}
\BIBentrySTDinterwordspacing

\bibitem{prefix}
W.~M. Y.~Z. SHENGLIN~ZHANG, YING~LIU, ``{PreFix: Switch Failure Prediction in
  Datacenter Networks},'' vol.~1, 2018.

\bibitem{YIN}
\BIBentryALTinterwordspacing
A.~D. Cheveigné and H.~Kawahara, ``{YIN, a fundamental frequency estimator for
  speech and music},'' vol.~1, 2002. [Online]. Available:
  \url{http://audition.ens.fr/adc/pdf/2002\_JASA\_YIN.pdf}
\BIBentrySTDinterwordspacing

\bibitem{pewma}
\BIBentryALTinterwordspacing
W.~W.~S. Kevin M.~Carter, ``{Probabilistic reasoning for streaming anomaly
  detection},'' 2012. [Online]. Available:
  \url{https://ieeexplore.ieee.org/document/6319708}
\BIBentrySTDinterwordspacing

\bibitem{kalman}
\BIBentryALTinterwordspacing
A.~Y. Aravkin, J.~V. Burke, and G.~Pillonetto, ``{Optimization viewpoint on
  Kalman smoothing, with applications to robust and sparse estimation.}'' 2013.
  [Online]. Available: \url{https://arxiv.org/pdf/1303.1993.pdf}
\BIBentrySTDinterwordspacing

\bibitem{dbscan}
\BIBentryALTinterwordspacing
H.~P. K. J.~S. Ester, M. and X.~Xu, ``{A Density-Based Algorithm for
  Discovering Clusters in Large Spatial Databases with Noise},'' pp. 226--231,
  1996. [Online]. Available:
  \url{https://www.aaai.org/Papers/KDD/1996/KDD96-037.pdf}
\BIBentrySTDinterwordspacing

\bibitem{kde}
D.~H. M.C.~Jonesa, ``{Maximum likelihood kernel density estimation: on the
  potential of convolution sieves}.''

\end{thebibliography}

\end{document}